# A Location-based Approach for Distributed Kiosk Design

Andy Luse
*Iowa State University*, andyluse@iastate.edu

Susana Berenice Vidrio-Baron
*Iowa State University*, sbvidrio@iastate.edu

Brian E. Mennecke
*Iowa State University*, mennecke@iastate.edu

Anthony M. Townsend
*Iowa State University*, amt@iastate.edu

# A Location-based Approach for Distributed Kiosk Design

**Andy Luse**
Iowa State University
andyluse@iastate.edu

**Susana Berenice Vidrio-Baron**
Iowa State University
sbvidrio@iastate.edu

**Brian E. Mennecke**
Iowa State University
mennecke@iastate.edu

**Anthony M. Townsend**
Iowa State University
amt@iastate.edu

**ABSTRACT**

Electronic kiosk interface design and implementation metrics have been well established. The problem arises when more than one kiosk is utilized in a different location within the same geographic proximity using the same basic informational parameters. This manuscript describes the design implications of a distributed kiosk environment from the standpoint of a field experiment. The log files from 2 kiosks deployed in the same building are analyzed for correlations among kiosk location and information required. The results show that while kiosk systems deployed in "primary entrances" should have a broad view of pertinent information, kiosks deployed in more remote locations should have information pertinent to that area initially presented to the individual. This research provides both confirmatory evidence and a checklist of implementation decision points for those who wish to implement a distributed kiosk architecture.

**INTRODUCTION**

Kiosk systems have become commonplace in today's technological society (Barab, Bowdish, Young, & Owen, 2004). Many different areas ranging from entertainment (Stapleton & Hughes, 2005) to healthcare (Nicholas, Huntington, & Williams, 2001b) utilize kiosk technology. The systems can consist of simple terminals designed to provide information and services to the general public (Maguire, 1999) to complex systems employing sound (Christian & Avery, 2000), user tracking through cameras (Christian & Avery, 1998), or speech recognition (Gauvain, Gangolf, & Lamel, 1996). The only common criteria is that kiosks are public-access terminals (Borchers, Deussen, & Knorzer, 1995).

A great deal of research has been done in the area of usability and evaluation both of user interfaces and more specifically kiosk systems. Kiosk systems are typically considered a subset of the wider area of user interface design (Shneiderman & Plaisant, 2004). The systems are first categorized on their use (Borchers et al., 1995) based on the intended audience of the system. Various metrics have also been employed to evaluate kiosk systems including log file examination (Nicholas et al., 2001b) and an overarching system for all aspects of kiosk design from placement to interface design (Maguire, 1999).

Location-based IS creates user experiences that are tailored to the physical proximity of the user and the system in use. This will result in a more precise information-centric view for the user that supports him or her by providing contextual information based on the location (Jose & Davies, 1999). This offers convenience because the user is presented with information which is more salient for his or her current location and removes the necessity for the user to make decisions about information relevant to their current situation.

A distributed kiosk environment employs multiple informational kiosk systems presenting similar information and utilized in differing location contexts within the same geographic proximity. These systems present a unique challenge both for informational needs and interface design. Distributed kiosks, if implemented correctly, can offer advantages for users by tailoring the interface to his or her current physical location. Furthermore, these systems can work together to offer different views of the same information that coincides with the current location of the kiosk/user.

The rest of this paper is organized as follows. The next section provides background information in kiosk design and evaluation as well as location-based information systems. Next, a distributed kiosk architecture is described. The following section describes the methods used in this study. Results and discussions are presented in the following two sections respectively. Finally, conclusions and future work are given.

## BACKGROUND

**Kiosk Design and Evaluation**

Kiosk systems come in a variety of types and are employed for a large array of uses. Kiosk categorization typically falls within one of 4 types (Borchers et al., 1995).

1. Informational: Provide information to the user within a limited subject field.
2. Advertising: Utilized by companies or institutions to present themselves and/or their products to the public.
3. Service: Like informational kiosks, with greater flow of information from the customer to the system.
4. Entertainment: Entertain the user.

This study incorporates an *Informational* kiosk design providing the user with information about people, rooms, and activities pertinent to the building in which the kiosks are located.

Various methods have been employed to evaluate kiosk/user interface design and usage. Hilbert and Redmiles perform a in-depth literature review to examine various techniques to extract usability information from user interface events (2000). A high-level overview of design principles for kiosks is presented by Kules et al from information gathered at the Computer Human Interaction (CHI) Conference in 2001 (2003). These principles include immediate attraction, immediate learning, immediate engagement, and immediate disengagement. Maguire also provides a detailed listing of design issues related to public information kiosk systems including location, user instructions for use, language, privacy, system input, display output, navigation, stakeholders, and testing (Maguire, 1999). Schneiderman also provides a number of guidelines regarding user interface design (2004) including consistency, feedback for actions, universal usability, use of dialog boxes, error prevention, reversibility of actions, internal locus of control, and short-term memory load reduction.

Log files are commonly used to keep track of system usage in various contexts. Websites keep extensive logs on user access including items viewed, time spent, pages viewed, etc. (Nicholas, 1996; Stout, 1997). Kiosk log files are also employed both as a metric of system usage and usability (Barab et al., 2004; Nicholas, Huntington, & Williams, 2001a; Nicholas et al., 2001b; Steiger & Suter, 1994). As with internet logs, these kiosk logs can keep track both of system use and usability for future improvements to the system. Metrics pertaining to kiosk usage can then be extracted from these logs using various statistical techniques to gain a greater understanding of the system and the user (Nicholas et al., 2001b).

**Location-based Information Systems**

Location-based information systems are built on the idea of connecting information pieces to locations in physical space (Persson, Espinoza, Fagerberg, Sandin, & Coster, 2003). Systems that employ location-based technologies utilize specific methods of data retrieval, including geographic queries and spatial queries. The term geographic queries and spatial queries imply querying a spatially indexed database based on relationships between particular items in that database within a particular geographic coordinate system. Spatial querying is the more general term. It can be defined as queries about the spatial relationships (intersection, containment, boundary, adjacency, proximity) of entries geometrically defined and located in space (Floriani, Marzano, & Puppo, 1993). Geographic querying assumes that the space is delineated by the well-defined coordinate systems of the "real world". An important component of geographic querying is to develop methods that can perform automatic geo referencing of information. By automatic geo referencing, we mean to automatically index and retrieve information according to the geographic locations discussed, displayed, or otherwise associated with its content (Larson, 1996).

Spatial relationships may be both geometric and topological (spatially related but without measureable distance or absolute direction). Some researchers suggest that there are two primary classes of requests from users: "What's here?" query and the "Where's this" query. The first type of query stems from a desire to discover what information is available about a particular location while the second stems from a desire to discover where certain phenomena occur (Frew et al., 1995).

**Information Foraging**

Access to information is more readily available today than ever before with the accessible nature of the Internet, but even with this availability humans must actively seek information they want or need. Information Foraging theory provides a framework in which to study the evolution of information seeking behavior (Pirolli & Card, 1999). The theory posits that, when feasible, natural information systems evolve. These can include human information systems as well as electronic information systems. Also, the primary problem in information gathering is attention allocation. Information foraging can provide a basis for user interface design to provide a more stimulating environment for users of systems while allowing for a small cognitive load.

**Distributed Kiosk Environment**

A distributed kiosk environment consists of multiple kiosks, typically in the same geographic region yet positioned at differing locations within that region, each designed to provide the same information. Though intended for the same task and informational needs, many times these kiosks will be employed for different subsets of tasks. As an example, let's consider a traditional poster-based shopping mall directory. Typically large shopping malls will have multiple directories located throughout the mall, each of which typically showing the exact same mall layout. Depending on where these directories are located, though, they may be employed for different uses. The directory at the main entrance may provide more of an overview mechanism where customers first come to find information about where they want to go in the mall. Another directory in a different location may primarily be utilized for finding information in that specific section of the mall. The shopper may have used the directory at the front doors to find where they wanted to go and then used the other directory to fine-tune their search once they got to this area. This is why most of these directories put a "you are here" marker on the directory. The important thing to notice is that both directories have exactly the same information, only the information used on each was very different.

The above analogy points to the need for fine-tuning the information presented to the user in different locations within the same geographic proximity. Also, the above research by Kules (2003) and Schneiderman (2004) provides guidelines for this research for gaining and retaining user attention as well as standardized usability metrics that should be considered to address user informational needs. Location-based IS research points to the need for information to have a link to the spatial context of the system. Also, research on information foraging reflects a need for information-based systems which accommodate a user's information seeking in their respective environment. A distributed kiosk environment aims to apply these principles. A kiosk in a particular location can have the display tailored to the search and wayfinding requirements associated with that context. But where the capabilities of the mall poster directory falls short is where the distributed electronic kiosk system excels. If, for example, a mall poster located distant from the main entrance only shows information about the proximity around the poster, then the user would be required to go back to the main entrance to find out information about places in the mall that are not shown on the local poster. Conversely, a distributed kiosk system can show information specific to the user's location but still allow the user to "step back" and see information about the entire geographic area. With the above background research in mind, we propose the following hypotheses.

> H1: The geographic location of the kiosk will be associated with the information the user is seeking.
> H1a: Utilization of a kiosk located at the main entrance(s) of a building will have a broad geographic focus of information retrieval
> H1b: Utilization of a kiosk located in a remote location of a geographic area will have a location-focused information retrieval pattern.

**METHODS**

The kiosks employed in this study provide information for visitors to a college of business building located on the central campus of a large Midwestern university. The kiosk is designed to give visitors and students information about rooms, classes, and activities within the college. The system allows the user to select heading categories of Faculty and Staff, Rooms, Named Spaces, Events, and Classes. After one of these primary headings is selected, the user is then presented with one or more list boxes containing information about the parent category. For example, when *Faculty/Staff Directory* is selected, two list boxes are presented to the user. The upper list box contains delineations of departments within the college, which allows the user to narrow the results in the second list box. The second list box contains listing information for the faculty and staff, listed by last name. When the user selects a faculty or staff member, a 2D map of the floor the faculty or staff member is located on comes into view and their respective room flashes in red. Also, relevant information pertaining to the selected individual is displayed in an information panel. This allows the user to easily locate the person they wish to find (see Figure 1).

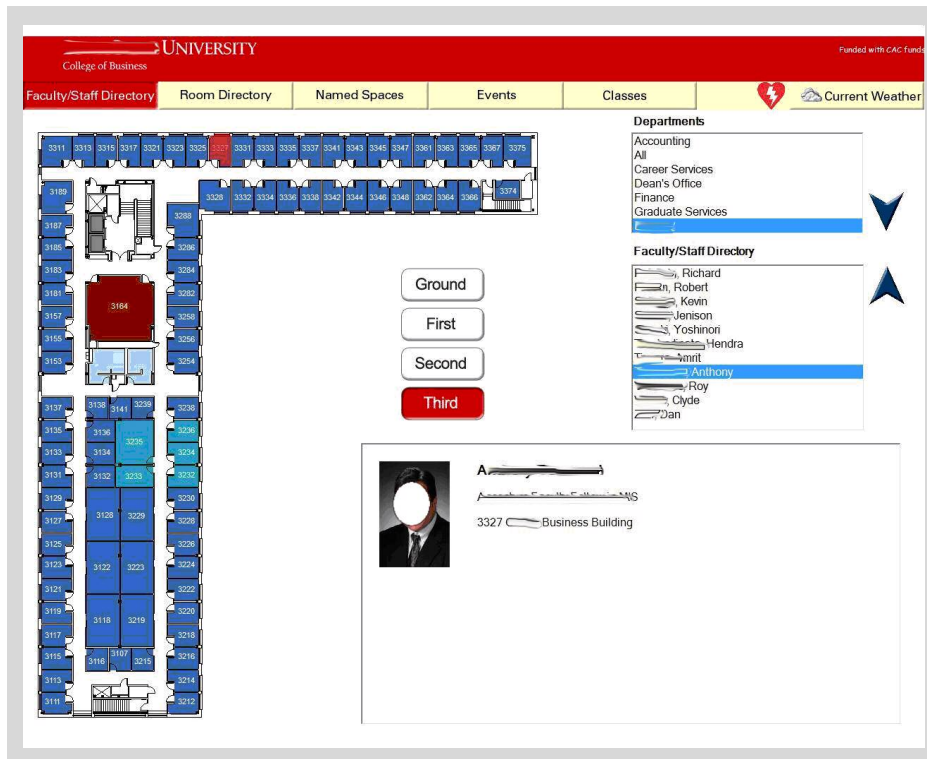

**Figure 1. Example Screen Capture from the Third Floor Kiosk.**

During kiosk design, a text file containing user click events was also designed to capture click behavior. This comma-separated file contains information for each finger touch on the kiosk including the x and y coordinates of the click on the screen, the name of the object clicked on the screen, the type of object clicked (list box, button, etc.), the room clicked (if a room was clicked), the floor present when the click occurred, and the time the click occurred. See Table 1 for an example subset from this file.

| x | y | objectClicked | typeObjectClicked | roomClicked | displayed/clicked floor | time |
|---|---|---|---|---|---|---|
| 136 | 103 | Faculty Staff Directory | Link | | none | 10/20/05 12:24 PM |
| 917 | 282 | Departments | Listbox | | none | 10/20/05 12:24 PM |
| 1015 | 455 | Faculty Staff Directory | Listbox | 3125 | Third | 10/20/05 12:24 PM |
| 1196 | 96 | Current Weather | Link | | Third | 10/20/05 12:24 PM |
| 1196 | 96 | Current Weather | Link | | Third | 10/20/05 1:36 PM |
| 490 | 103 | Named Spaces | Link | | Third | 10/20/05 1:37 PM |
| 1032 | 252 | Named Spaces | Listbox | 2140 | Second | 10/20/05 1:37 PM |
| 1035 | 375 | Named Spaces | Listbox | 1145 | First | 10/20/05 1:37 PM |
| 979 | 448 | Named Spaces | Listbox | 1148 | First | 10/20/05 1:37 PM |
| 1002 | 518 | Named Spaces | Listbox | 2113 | Second | 10/20/05 1:37 PM |
| 1035 | 335 | Named Spaces | Listbox | 2134 | Second | 10/24/05 11:34 AM |
| 1193 | 110 | Named Spaces | Listbox | 1140 | First | 10/24/05 11:34 AM |

**Table 1. Subset of Kiosk Logs**

Two electronic kiosks are used in the building. Both electronic kiosks are located in an area proximate to the elevators; one is on the first floor and one on the third floor of the building. There are 4 floors total in the building including ground, first, second, and third. The same software is employed on both kiosks with users on each floor seeing exactly the same information displayed in exactly the same way. Information is updated nightly by syncing with the College's database server. The kiosks were deployed around October, 2005. Analysis was performed on the logs for each with a cutoff date of November 2007 (around 2 years of log data for each).

**RESULTS**

In total, there were 64,218 click events analyzed for the first floor kiosk and 32,152 for the third floor kiosk. The study examined one independent variable and four dependent variables. The independent variable consists of the physical kiosk location while the dependent variable consists of the displayed/clicked floor information presented on the screen to the user after his or her selection. A MANOVA test was used to analyze the data for significant variances among the dependent and independent variables. For the purposes of the test, the independent variable was coded as a dichotomous variable based on the kiosk location (1 representing the first floor 2 for the third floor). The dependent variables were coded as four unique binary variables with each representing a given floor (i.e., ground, first, second, or third). For each floor, the binary variable was coded as 1 if the user clicked that floor and 0 if the user clicked another floor. The MANOVA was used to analyze the relationship between the kiosk used and floor selected in one omnibus model which accounted for the interdependence between each of the dependent variables. Normality tests for each separate dependent variable were run. While the Lilliefors test for normality showed each variable to be from a non-normal distribution, the large sample size will allow the violations to have little impact on the overall MANOVA (Hair, Anderson, Tatham, & Black, 1998).

The overall $R^2$ of the model is 0.13 which means we are explaining 13% of the variance among the variables with a significance of < 0.001. The results from the MANOVA analysis can be seen in Table 2 below.

**Tests of Between-Subjects Effects**

| Source | Dependent Variable | Type III Sum of Squares | df | Mean Square | F | Sig. |
|---|---|---|---|---|---|---|
| Corrected Model | Clicked Ground | 282.502[a] | 1 | 282.502 | 2202.826 | .000 |
| | Clicked First Floor | 933.376[b] | 1 | 933.376 | 5628.926 | .000 |
| | Clicked Second Floor | 180.448[c] | 1 | 180.448 | 1311.723 | .000 |
| | Clicked Third Floor | 3695.675[d] | 1 | 3695.675 | 17704.062 | .000 |
| Intercept | Clicked Ground | 1581.923 | 1 | 1581.923 | 12335.154 | .000 |
| | Clicked First Floor | 3161.079 | 1 | 3161.079 | 19063.570 | .000 |
| | Clicked Second Floor | 1974.616 | 1 | 1974.616 | 14354.029 | .000 |
| | Clicked Third Floor | 22864.918 | 1 | 22864.918 | 109534.0 | .000 |
| kiosk | Clicked Ground | 282.502 | 1 | 282.502 | 2202.826 | .000 |
| | Clicked First Floor | 933.376 | 1 | 933.376 | 5628.926 | .000 |
| | Clicked Second Floor | 180.448 | 1 | 180.448 | 1311.723 | .000 |
| | Clicked Third Floor | 3695.675 | 1 | 3695.675 | 17704.062 | .000 |
| Error | Clicked Ground | 12231.246 | 95374 | .128 | | |
| | Clicked First Floor | 15814.706 | 95374 | .166 | | |
| | Clicked Second Floor | 13120.150 | 95374 | .138 | | |
| | Clicked Third Floor | 19909.064 | 95374 | .209 | | |
| Total | Clicked Ground | 14815.000 | 95376 | | | |
| | Clicked First Floor | 21672.000 | 95376 | | | |
| | Clicked Second Floor | 15977.000 | 95376 | | | |
| | Clicked Third Floor | 42911.000 | 95376 | | | |
| Corrected Total | Clicked Ground | 12513.748 | 95375 | | | |
| | Clicked First Floor | 16748.083 | 95375 | | | |
| | Clicked Second Floor | 13300.598 | 95375 | | | |
| | Clicked Third Floor | 23604.739 | 95375 | | | |

a. R Squared = .023 (Adjusted R Squared = .023)
b. R Squared = .056 (Adjusted R Squared = .056)
c. R Squared = .014 (Adjusted R Squared = .014)
d. R Squared = .157 (Adjusted R Squared = .157)

**Table 2. MANOVA Results for Analysis of Electronic Kiosk Click Behavior**

In table 3 the descriptive statistics for the analysis are shown. The table shows that the probability of the user requesting information regarding the third floor when located in front of the third floor kiosk is 73% which is much greater than the other probabilities shown.

**Descriptive Statistics**

|  | kiosk | Mean | Std. Deviation | N |
|---|---|---|---|---|
| Clicked Ground | 1 | .1940 | .39544 | 63374 |
|  | 3 | .0787 | .26934 | 32002 |
|  | Total | .1553 | .36222 | 95376 |
| Clicked First Floor | 1 | .2975 | .45718 | 63374 |
|  | 3 | .0880 | .28334 | 32002 |
|  | Total | .2272 | .41905 | 95376 |
| Clicked Second Floor | 1 | .1984 | .39882 | 63374 |
|  | 3 | .1063 | .30823 | 32002 |
|  | Total | .1675 | .37344 | 95376 |
| Clicked Third Floor | 1 | .3100 | .46251 | 63374 |
|  | 3 | .7269 | .44555 | 32002 |
|  | Total | .4499 | .49749 | 95376 |

**Table 3. Descriptive Statistics for Electronic Kiosk Click Behavior**

To verify and validate the above results, a cluster analysis was also performed on the logs for each of the 2 kiosks separately. The Expectation Maximization (EM) algorithm was employed with a node limit of 10. The algorithm found that the overall probability of all clusters in the model has a probability of 0.72 for the third floor being present for a click event that occurred on the third floor kiosk. The same analysis was then run on the logs from the first floor kiosk. The algorithm found that the overall probability of all clusters found a similar probability ranking across all floors with 0.19, 0.30, 0.20, and 0.30 respectively.

Both of the above analyses provide strong support for the research hypotheses. For the omnibus hypothesis, Hypothesis H1, the results from the MANOVA show that there is a significant relationship between physical kiosk location and user informational needs. Also, the descriptive statistics and the cluster analysis support Hypotheses H1a and H1b and show that there is a high probability of an electronic kiosk located on the third floor (a specific remote location) being used to conduct information searches focused on the area proximate to the kiosk. On the other hand, the focus of searches for the kiosk on the first floor (i.e., the main entrance) was more evenly distributed among the floors in the building.

**DISCUSSION**

The objective of this research was to examine the relationship between kiosk location and user informational searches. We hypothesized that (1) there would be a significant relationship between the physical kiosk location and user searches and (2) there would be a higher proximity-based user search pattern for the kiosk located in a remote location in the building. The analyses confirm our hypotheses.

The primary contribution of this research is in the area of kiosk interface design for a distributed kiosk environment. This provides researchers and practitioners with design strategies for location-based information architectures. Specifically, we show that the informational needs of users of a kiosk located at the main entrance of a building are broader in nature and thus a more general overview is needed for display. In addition, informational needs of users of a kiosk located in a specific area are more directed towards information specific to the physical location of the kiosk. While the results of this research may appear to be obvious, this is the first and only study completed to date that has examined the pattern of usage of electronic kiosks given location and task-related considerations. Given this, plus the large and representative sample of data used for this analysis, this study provides a sound basis for presenting information sources to users in uncontrolled public spaces. Based on these results, we offer a checklist for designers who may wish to implement a user-centered distributed kiosk system (see Table 4).

| Category | Decision points |
|---|---|
| Kiosk availability | money |
|  | kiosk location needs |
| Kiosk Location | user informational needs |
|  | structure of geographic area |
| Kiosk display information | interface size |
|  | kiosk location |
|  | kiosk location type |
|  | user informational needs |

**Table 4. Checklist for Distributed Kiosk Implementation.**

The above decision points provide highlights when implementing a distributed kiosk environment. In order to provide a more comprehensive listing, both points derived from the above research as well as standard metrics for kiosk design are included. Kiosk availability relates to the number of kiosks in the geographic area of the distributed kiosk system. The number of kiosks used in the area depends both on financial resources (Carr, 2003) as well as the needs of the users for number of kiosk locations. Kiosk location describes the physical location of each kiosk after it has been purchased. This again depends on user needs within the geographic area as well as the structure of the area intended for a distributed kiosk environment. Kiosk display information describes the content used for display on each kiosk once it has placed. Interface size restricts the type of information display (Borchers et al., 1995) and typically is a financially-driven decision. The above research also has shown that the kiosk location, and location type, as well as user information needs within the location, are important decision points in an overall distributed kiosk environment.

**LIMITATIONS AND FUTURE WORK**

This research offers valuable insights into the relationship between kiosks in a distributed kiosk environment and how these kiosks can be made more effective based on their location. Nevertheless, there are a few limitations to the study. First, kiosks were only located on the first and third floor and not on the ground or second floor. Second, both kiosks were placed in the same area on both floors, namely proximate to the elevators, which limit our ability to speculate about the role that proximity may have on other important features in the building. Despite these limitations, the research provides verification of a significant correlation between physical kiosk location and user informational needs. Furthermore, a checklist provides guidance for those interested in implementing a distributed kiosk environment.

Future directions may include implementing a distributed kiosk system with more than two physical kiosks. Also, future research could address questions that might arise when multiple kiosks are placed on the same building floor but in different physical locations. Additional research may also consider display design layout issues based on user experimental studies.

**ACKNOWLEDGMENTS**

This research was made possible by the Iowa State University College of Business with the support of Greg Buttery.

**REFERENCES**

1. Barab, S. A., Bowdish, B. E., Young, M. F., and Owen, S. V. (2004) Understanding kiosk navigation: Using log files to capture hypermedia searches. *Instructional Science,* 24, 5, 377-395.
2. Borchers, J., Deussen, O., and Knorzer, C. (1995) Getting it across: layout issues for kiosk systems. *ACM SIGCHI Bulletin,* 27, 4, 68-74.
3. Carr, N. G. (2003) IT Doesn't Matter. *Harvard Business Review,* 81, 5, 41-49.
4. Christian, A. D., and Avery, B. L. (1998) Digital smart kiosk project, *Proceedings of the SIGCHI conference on Human factors in computing systems* Los Angeles, California, United States, ACM Press/Addison-Wesley Publishing Co., 155 - 162.

5. Christian, A. D., and Avery, B. L. (2000) Speak out and annoy someone: experience with intelligent kiosks, *Proceedings of the SIGCHI conference on Human factors in computing systems* The Hague, The Netherlands, ACM, 313 - 320.
6. Floriani, L. D., Marzano, P., and Puppo, E. (1993) Spatial queries and data models. *Lecture Notes in Computer Science,* 716, 113-138.
7. Frew, J., Carver, L., Fischer, C., Goodchild, M., Larsgaard, M., Smith, T., et al. (1995) The Alexandria rapid prototype: building a digital library for spatial information, *ESRI User Conference* Environmental Systems Research Institute, Inc., Redlands, CA.
8. Gauvain, J. L., Gangolf, J. J., and Lamel, L. (1996) Speech recognition for an information kiosk, *Proceedings of the Fourth International Conference on Spoken Language, 1996. ICSLP 96* Philadelphia, PA, USA, IEEE, 849-852.
9. Hair, J. F., Anderson, R. E., Tatham, R. L., and Black, W. C. (1998) Multivariate data analysis (5th ed.), Prentice-Hall, Upper Saddle River, NJ.
10. Hilbert, D. M., and Redmiles, D. F. (2000) Extracting usability information from user interface events. *ACM Computing Surveys,* 32, 4, 384 - 421.
11. Jose, R., and Davies, N. (1999) Scalable and flexible location-based services for ubiquitous information access. *Lecture Notes in Computer Science,* 1707, 52-66.
12. Kules, B., Kang, H., Plaisant, C., Rose, A., and Shneiderman, B. (2003). Immediate usability: kiosk design principles from the chi 2001 photo library [electronic version]. *Tech. Report CS-TR-4293*, U. Maryland,
13. Larson, R. (1996) Geographic information retrieval and spatial browsing, *1995 Clinic on Library Applications of Data Processing*, Graduate School of Library and Information Science, University of Illinois at Urbana-Champaign, 81-124.
14. Maguire, M. C. (1999) Review of user-interface design guidelines for public information kiosk systems. *International Journal of Human-Computers Studies,* 50, 3, 263-286.
15. Nicholas, D. (1996) An assessment of the online searching behavior of practitioner end users. *Journal of Documentation,* 52, 3, 227-251.
16. Nicholas, D., Huntington, P., and Williams, P. (2001a) Comparing web and touch screen transaction log files. *Journal of Medical Internet Research,* 3, 2.
17. Nicholas, D., Huntington, P., and Williams, P. (2001b) Establishing metrics for the evaluation of touch screen kiosks. *Journal of Information Science,* 27, 2, 61-71.
18. Persson, P., Espinoza, F., Fagerberg, P., Sandin, A., and Coster, R. (2003). GeoNotes: a location-based information system for public spaces. In *Designing information spaces: the social navigation approach* (pp. 151-173). London, UK: Springer-Verlag.
19. Pirolli, P., and Card, S. K. (1999) Information foraging. *Psychological Review,* 106, 4, 643-675.
20. Shneiderman, B., and Plaisant, C. (2004) Designing the user interface: Strategies for effective human-computer interaction (4th ed.), Pearson/Addison Wesley.
21. Stapleton, C. B., and Hughes, C. E. (2005) Mixed reality and experiential movie trailers: Combining emotions and immersion to innovate entertainment marketing, *Proceedings of the 2005 International Conference on Human–Computer Interface Advances in Modeling and Simulation*, 40-48.
22. Steiger, P., and Suter, B. A. (1994) MINNELLI—Experiences with an interactive information kiosk for casual users, *Proceedings of the UBILAB Conference '94* Zurich, Switzlerand, 124-133.
23. Stout, R. (1997) Web site stats: Tracking hits and analyzing traffic, Osborne McGraw-Hill.